\documentclass[a4paper,11pt]{article}

\usepackage{jheppub} 
\usepackage{amsmath,amssymb,amsthm,bm,bigdelim,multirow}
\usepackage{color}
\usepackage{hyperref}
\usepackage{dsfont}

\newcommand{\kslash}{k\kern-1ex /}
\newcommand{\pslash}{p\kern-1ex /}
\newcommand{\qslash}{q\kern-1ex /}
\newcommand{\lslash}{l\kern-1ex /}
\newcommand{\sslash}{s\kern-1ex /}
\newcommand{\Dslash}{D\kern-1.2ex /}

\newcommand{\beqa}{\begin{eqnarray}}
\newcommand{\eeqa}{\end{eqnarray}}

\newcommand{\bd}{\begin{description}}
\newcommand{\ed}{\end{description}}
\newcommand{\la}{\langle}
\newcommand{\ra}{\rangle}
\newcommand{\ben}{\begin{eqnarray}}
\newcommand{\een}{\end{eqnarray}}

\def\lsim{\raise0.3ex\hbox{$<$\kern-0.75em\raise-1.1ex\hbox{$\sim$}}}
\def\gsim{\raise0.3ex\hbox{$>$\kern-0.75em\raise-1.1ex\hbox{$\sim$}}}
\def\simgt{\rlap{\lower 3.5 pt\hbox{$\mathchar \sim$}}\raise 2.0pt \hbox {$>$}}
\def\simlt{\rlap{\lower 3.5 pt\hbox{$\mathchar \sim$}}\raise 2.0pt \hbox {$<$}}

\begin{document}

\title{Critical endpoint of (3+1)-dimensional finite density $\mathds{Z}_3$ gauge-Higgs model with tensor renormalization group}

\author[a]{Shinichiro Akiyama}
	\affiliation[a]{Institute for Physics of Intelligence, University of Tokyo, Tokyo, 113-003, Japan}
    	\emailAdd{akiyama@phys.s.u-tokyo.ac.jp}

  	\author[b]{Yoshinobu Kuramashi}
  	\affiliation[b]{Center for Computational Sciences, University of Tsukuba, Tsukuba, Ibaraki
    305-8577, Japan}
  	\emailAdd{kuramasi@het.ph.tsukuba.ac.jp}

\abstract{
The critical endpoint of the (3+1)-dimensional $\mathds{Z}_3$ gauge-Higgs model at finite density is determined by the tensor renormalization group method. 
This work is an extension of the previous one on the $\mathds{Z}_2$ model. 
The vital difference between them is that the $\mathds{Z}_3$ model suffers from the sign problem, while the $\mathds{Z}_2$ model does not. 
We show that the tensor renormalization group method allows us to locate the critical endpoint for the $\mathds{Z}_3$ gauge-Higgs model at finite density, regardless of the sign problem.
}
\date{\today}

\preprint{UTHEP-780, UTCCS-P-147}

\maketitle

\section{Introduction}
\label{sec:intro}

The last decade was devoted to an initial stage to apply the tensor renormalization group (TRG) method \footnote{In this paper, the ``TRG method" or the ``TRG approach" refers to not only the original numerical algorithm proposed by Levin and Nave \cite{Levin:2006jai} but also its extensions \cite{Gu:2010yh,PhysRevB.86.045139,Shimizu:2014uva,Sakai:2017jwp,Adachi:2019paf,Kadoh:2019kqk,Akiyama:2020soe,PhysRevB.105.L060402,Kadoh:2021fri}.}, which was originally proposed to study two-dimensional (2$d$) classical spin systems in the field of condensed matter physics~\cite{Levin:2006jai}, to quantum field theories consisting of scalar, fermion, and gauge fields~\cite{Banuls:2019rao,Meurice:2020pxc,Okunishi:2021but}. 
There were many attempts to confirm or utilize the following expected advantages of the TRG method employing the lower-dimensional models: 
(i) no sign problem \cite{Denbleyker:2013bea,Shimizu:2014uva,Shimizu:2014fsa,Takeda:2014vwa,Kawauchi:2016xng,Shimizu:2017onf,Kadoh:2018hqq,Kadoh:2019ube,Kuramashi:2019cgs,Butt:2019uul,Takeda:2021mnc,Nakayama:2021iyp}, 
(ii) logarithmic computational cost on the system size, 
(iii) direct manipulation of the Grassmann variables \cite{Gu:2010yh,Shimizu:2014uva,Takeda:2014vwa,Sakai:2017jwp,Yoshimura:2017jpk,Kadoh:2018hqq,Akiyama:2021xxr,Bloch:2022vqz}, 
(iv) evaluation of the partition function or the path-integral itself. 

\begin{figure}[htbp]
	\centering
	\includegraphics[width=0.9\hsize]{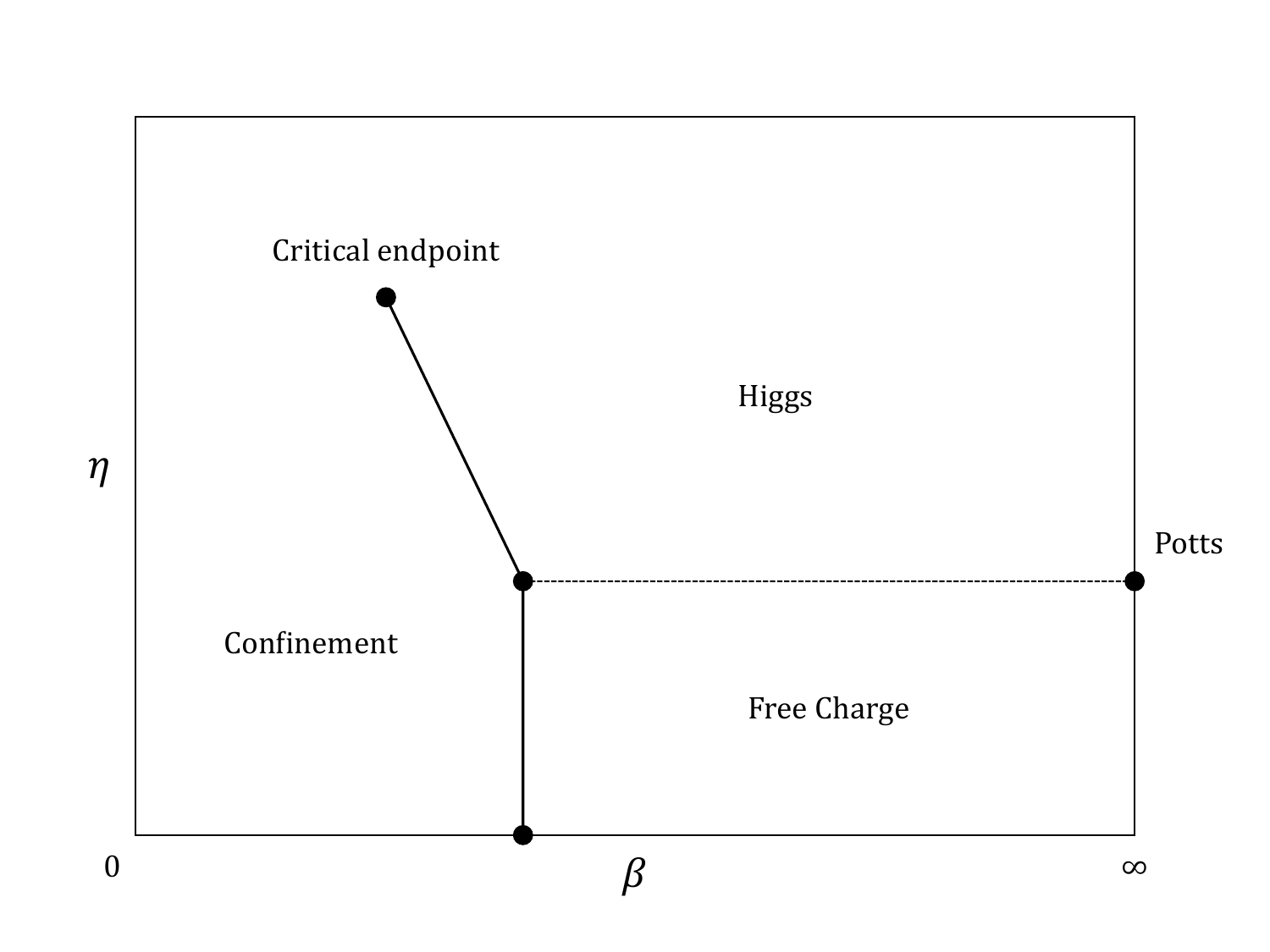}
 	\caption{Schematic phase diagram of (3+1)$d$ $\mathds{Z}_3$ gauge-Higgs model at the vanishing chemical potential. $\beta$-axis denotes the inverse gauge coupling and $\eta$ represents the spin-spin coupling.
          The pure $\mathds{Z}_{3}$ gauge theory is characterized by $\eta=0$ whose transition point is $\beta=2\ln(1+\sqrt{3})/3$~\cite{Balian:1974ts,Balian:1974ir,Balian:1974xw,KorthalsAltes:1978tp,Yoneya:1978dt}.
          The limit $\beta\to\infty$ is equivalent to the three-state Potts model whose transition point is obtained by the Monte Carlo renormalization group method~\cite{PhysRevLett.43.799,Caracciolo:1986ik}, where the estimated transition point of $4d$ Potts model is $2\eta=0.3875$.}
  	\label{fig:phasedgm}
\end{figure}

The first TRG calculation of the 4$d$ Ising model~\cite{Akiyama:2019xzy} was the trigger to explore various (3+1)$d$ quantum field theories with the TRG method: complex $\phi^4$ theory at finite density~\cite{Akiyama:2020ntf}, real $\phi^4$ theory~\cite{Akiyama:2021zhf}, Nambu$-$Jona-Lasinio (NJL) model at high density and very low temperature~\cite{Akiyama:2020soe}, and $U(N)$ gauge theory with the infinite-coupling limit~\cite{Milde:2021vln}. 
Recently, the phase structure of the $\mathds{Z}_2$ gauge-Higgs model at finite density has been investigated and its critical endpoint has been determined within the TRG method~\cite{Akiyama:2022eip}. 
This is the first application of the TRG method to a (3+1)$d$ lattice gauge theory beyond the (1+1)$d$ systems~\cite{Shimizu:2014uva,Shimizu:2014fsa,Shimizu:2017onf,Unmuth-Yockey:2018ugm,Kuramashi:2019cgs,Bazavov:2019qih,Fukuma:2021cni,Hirasawa:2021qvh} and (2+1)$d$ ones~\cite{Dittrich:2014mxa,Kuramashi:2018mmi,Unmuth-Yockey:2018xak}. 
In this paper, we investigate the phase structure of the $\mathds{Z}_3$ gauge-Higgs model at finite density.  Figure~\ref{fig:phasedgm} illustrates the expected phase diagram at vanishing density, which is speculated from the numerical study of the phase diagrams of the $\mathds{Z}_{N}$ gauge-Higgs models~\cite{Creutz:1979he,Creutz:1983ev,Baig:1987ka}. 
We determine the critical endpoint following the procedure employed in the $\mathds{Z}_2$ study~\cite{Akiyama:2022eip}. 
The important difference between the $\mathds{Z}_3$ and $\mathds{Z}_2$ models is that the former yields the sign problem at finite chemical potential contrary to the latter. 
Therefore, this model has been investigated by dual lattice simulations~\cite{Gattringer:2012jt,Langfeld:2018ykv}.
The purpose of this work is to confirm the effectiveness of the TRG method for a (3+1)$d$ gauge theory with the sign problem, which should serve as a better test bed for the future study of the QCD at finite density.    

This paper is organized as follows. 
In Sec.~\ref{sec:method}, we define the $\mathds{Z}_3$ gauge-Higgs model at finite density on a (3+1)$d$ lattice. 
In Sec.~\ref{sec:results}, we provide a consistency check between the TRG approach and the dual lattice simulations before we determine the critical endpoints at $\mu=0$, $1$, $2$ in the (3+1)$d$ model and discuss to what extent they are shifted by the effect of finite $\mu$. 
Section~\ref{sec:summary} is devoted to summary and outlook.

\section{Formulation and numerical algorithm}
\label{sec:method}

We consider the path integral of the $\mathds{Z}_3$ gauge-Higgs model at finite density on an isotropic hypercubic lattice $\Lambda_{3+1}=\{(n_1,\dots,n_{4})\ \vert n_{\nu}=1,\dots ,L\}$ whose volume is equal to $V=L^{4}$. The lattice spacing $a$ is set to $a=1$ without loss of generality. 
The gauge fields $U_{\nu}(n)$ ($\nu=1,\dots,4$) reside on the links and the matter fields $\sigma(n)$ are on the sites. 
Both variables $U_{\nu}(n)$ and $\sigma(n)$ take their values on $\mathds{Z}_{3}=\{1,\exp\left({\rm i}\pi/3\right),\exp\left({\rm i}2\pi/3\right)\}$.
The action $S$ is defined as
\begin{align}
\label{eq:action}
	S=
	&-{\beta}\sum_{n\in\Lambda_{3+1}}\sum_{\nu>\rho}
	\Re\left[U_{\nu}(n)U_{\rho}(n+\hat{\nu})U^{*}_{\nu}(n+\hat{\rho})U^{*}_{\rho}(n)\right]
	\nonumber\\
	&-\eta\sum_{n}\sum_{\nu}
	\left[
	{\rm e}^{\mu\delta_{\nu,4}}\sigma^{*}(n) U_{\nu}(n) \sigma(n+{\hat\nu})
	+{\rm e}^{-\mu\delta_{\nu,4}}\sigma^{*}(n) U^{*}_{\nu}(n-{\hat \nu})\sigma(n-{\hat\nu})
	\right],
\end{align}
where 
$\beta$ is the inverse gauge coupling, $\eta$ is the spin-spin coupling and $\mu$ is the chemical potential. 
This parametrization follows Ref.~\cite{Gattringer:2012jt}. 
We employ periodic boundary conditions for both the gauge and matter fields in all directions.
The path integral is then given by 
\begin{align}
\label{eq:Z}
	Z=\left(\prod_{n,\nu}\sum_{U_{\nu}(n)\in \mathds{Z}_{3}}\right)\left(\prod_{n}\sum_{\sigma(n)\in \mathds{Z}_{3}}\right){\rm e}^{-S},
\end{align}
where the sum is taken over all possible field configurations. Since $\sigma(n)\in\mathds{Z}_{3}$, one is allowed to choose the so-called unitary gauge \cite{Creutz:1979he}, which eliminates the matter field $\sigma(n)$ by redefining the link variable $U_{\nu}(n)$ via
\begin{align}
	\sigma^{*}(n) U_{\nu}(n) \sigma(n+{\hat\nu})\mapsto U_{\nu}(n).
\end{align}
With the unitary gauge, Eq.~\eqref{eq:action} is reduced to be
\begin{align}
	S=
	&-{\beta}\sum_{n\in\Lambda_{3+1}}\sum_{\nu>\rho}
	\Re\left[U_{\nu}(n)U_{\rho}(n+\hat{\nu})U^{*}_{\nu}(n+\hat{\rho})U^{*}_{\rho}(n)\right]
	\nonumber\\
	&-2\eta\sum_{n}\sum_{\nu}
	\left\{
	\cosh\left(\mu\delta_{\nu,4}\right)\Re\left[U_{\nu}(n)\right]
	+{\rm i}\sinh\left(\mu\delta_{\nu,4}\right)\Im\left[U_{\nu}(n)\right]
	\right\},
\end{align}
whose path integral is given by
\begin{align}	
\label{eq:Z_ug}
	Z=\left(\prod_{n,\nu}\sum_{U_{\nu}(n)\in \mathds{Z}_{3}}\right){\rm e}^{-S},
\end{align}
instead of Eq.~\eqref{eq:Z}.
The construction of the tensor network representation for the general $\mathds{Z}_{N}$ case is already described in Ref.~\cite{Akiyama:2022eip}, which is based on the asymmetric construction established in Ref.~\cite{Liu:2013nsa}. 
We also employ the anisotropic TRG (ATRG)~\cite{Adachi:2019paf} with the parallel computation as in Refs.~\cite{Akiyama:2022eip,Akiyama:2020ntf}.

\section{Numerical results} 
\label{sec:results}

\begin{figure}[htbp]
	\centering
	\includegraphics[width=0.75\hsize]{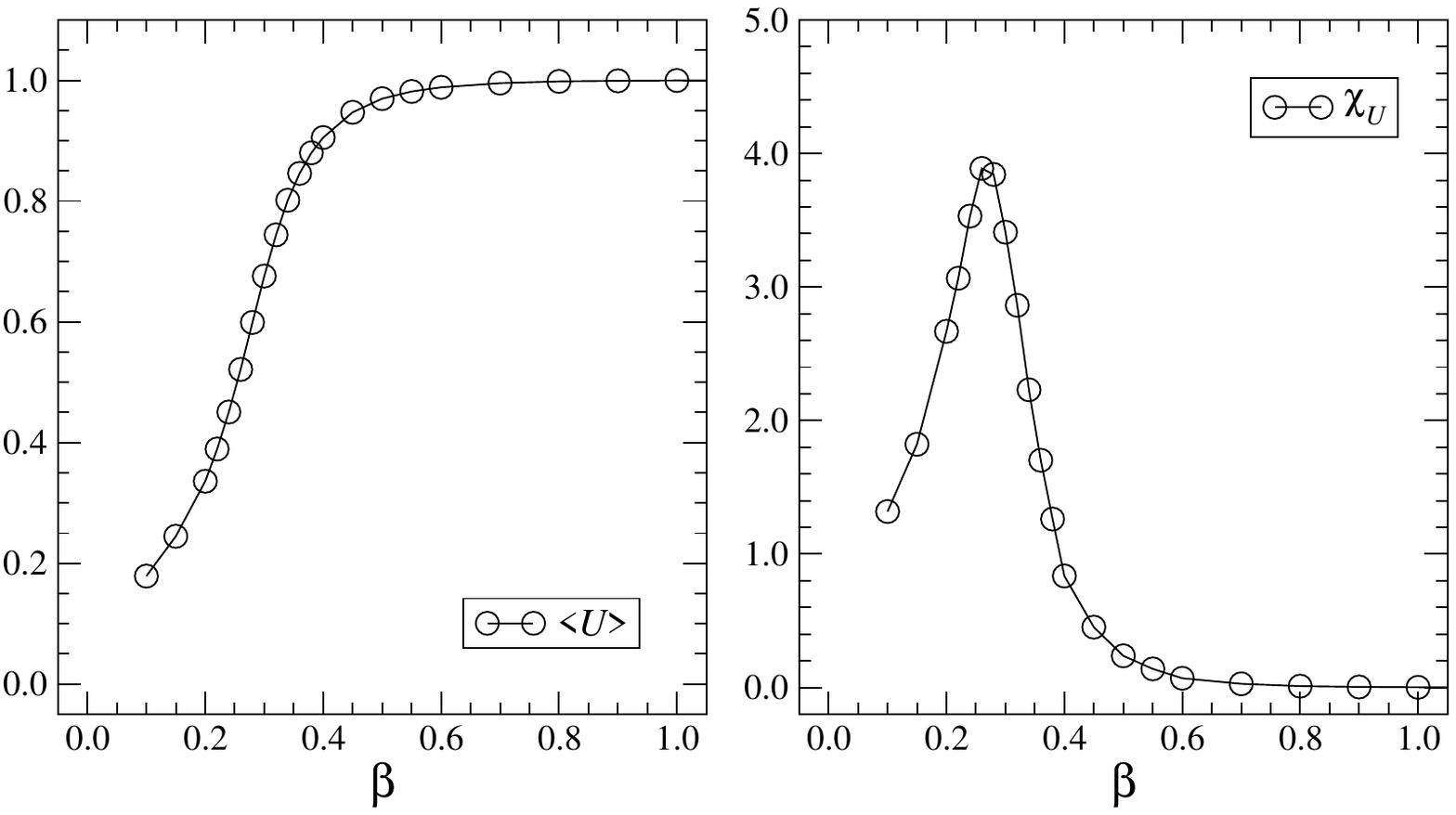}
	\caption{
	$\beta$ dependence of the plaquette value (left) and its susceptibility (right) at $\eta=0.5$ with $D=45$.
	}
  	\label{fig:plaq}
\end{figure}

The path integral of Eq.~\eqref{eq:Z_ug} is evaluated using the parallelized ATRG algorithm with the bond dimension $D$.
Firstly, we make a consistency check between our results and those obtained by the previous lattice calculation. 
In Ref.~\cite{Gattringer:2012jt}, the dual lattice simulation gives $\beta$ dependence of the plaquette value and its susceptibility choosing $\eta=0.5$,
where no phase transition is expected passing by the critical endpoint in Fig.~\ref{fig:phasedgm_mu} below. 
We evaluate the plaquette value $\langle U\rangle$ defined by
\begin{align}
\label{eq:number}
	\langle U\rangle=\frac{1}{6V}\frac{\partial\ln Z}{\partial \beta}
\end{align}
using the impurity tensor method, setting $D=45$.
To evaluate $\langle U\rangle$ via the impurity tensor method, we construct the tensor network representation differently from Ref.~\cite{Akiyama:2022eip}. 
The detail is explained in Appendix~\ref{app:plaq}.
In Fig.~\ref{fig:plaq}, we plot the plaquette value and its susceptibility $\chi_{U}$ as a function of $\beta$ at $\eta=0.5$. 
We calculate $\chi_{U}$ via the forward difference of $\langle U\rangle$.
The results should be compared with Fig.~5 in Ref.~\cite{Gattringer:2012jt} obtained with the dual lattice simulation. 
Both results show consistency with Ref.~\cite{Gattringer:2012jt}, where no sign of the phase transition is observed. 
Therefore, we expect that $D=45$ is sufficiently large to calculate the thermodynamic quantities in the finite-$\eta$ regime.
Note that the peak height of $\chi_{U}$ is a little bit smaller than that in Ref.~\cite{Gattringer:2012jt}.
We think that this can be attributed to the forward difference of $\langle U\rangle$.

\begin{figure}[htbp]
	\centering
	\includegraphics[width=0.75\hsize]{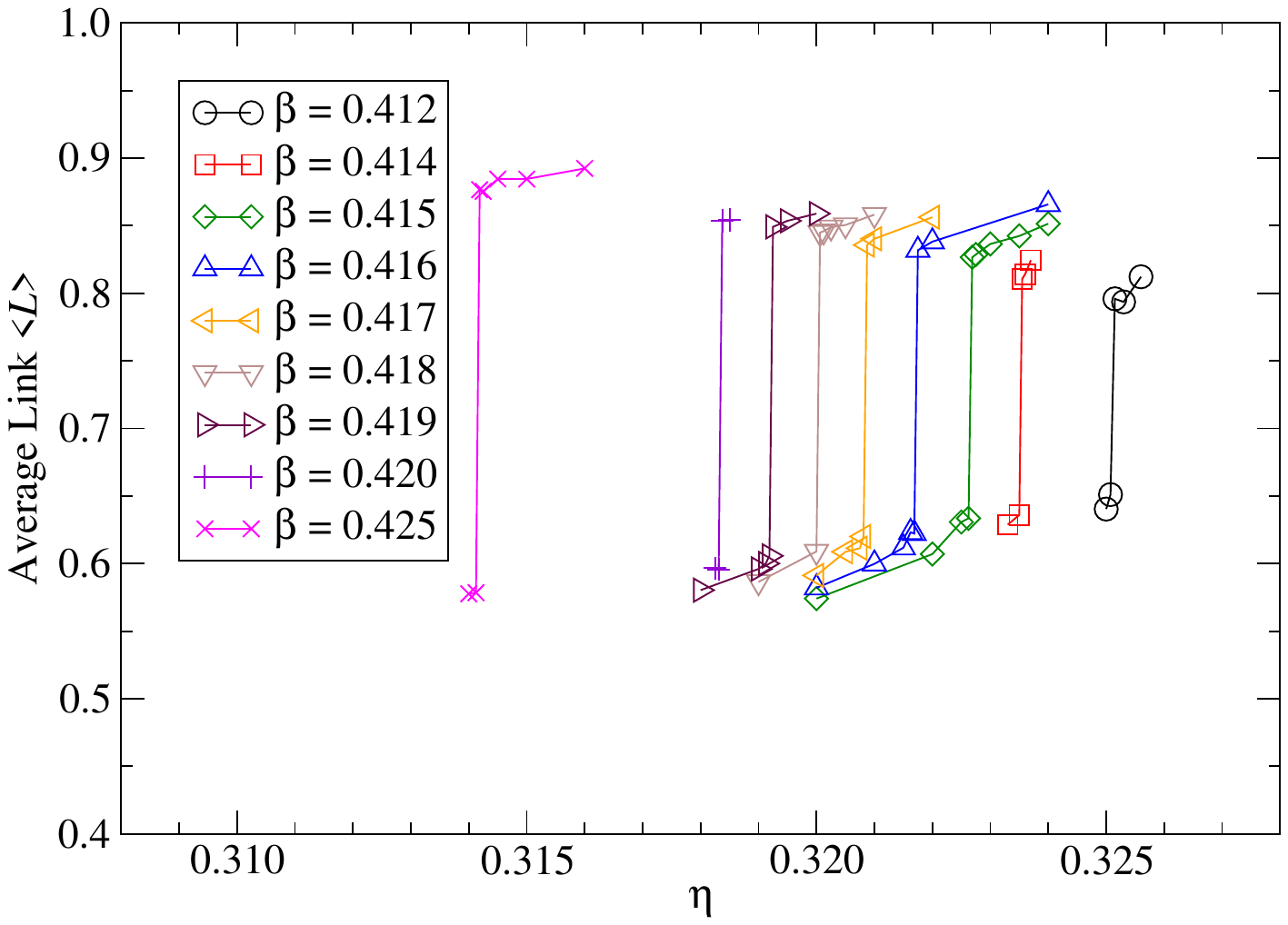}
	\caption{$\eta$ dependence of $\langle L\rangle$ at $\mu=0$ for $\beta\in[0.412,0.425]$.}
  	\label{fig:link_4d_mu=0}
\end{figure}

\begin{figure}[htbp]
	\centering
	\includegraphics[width=0.9\hsize]{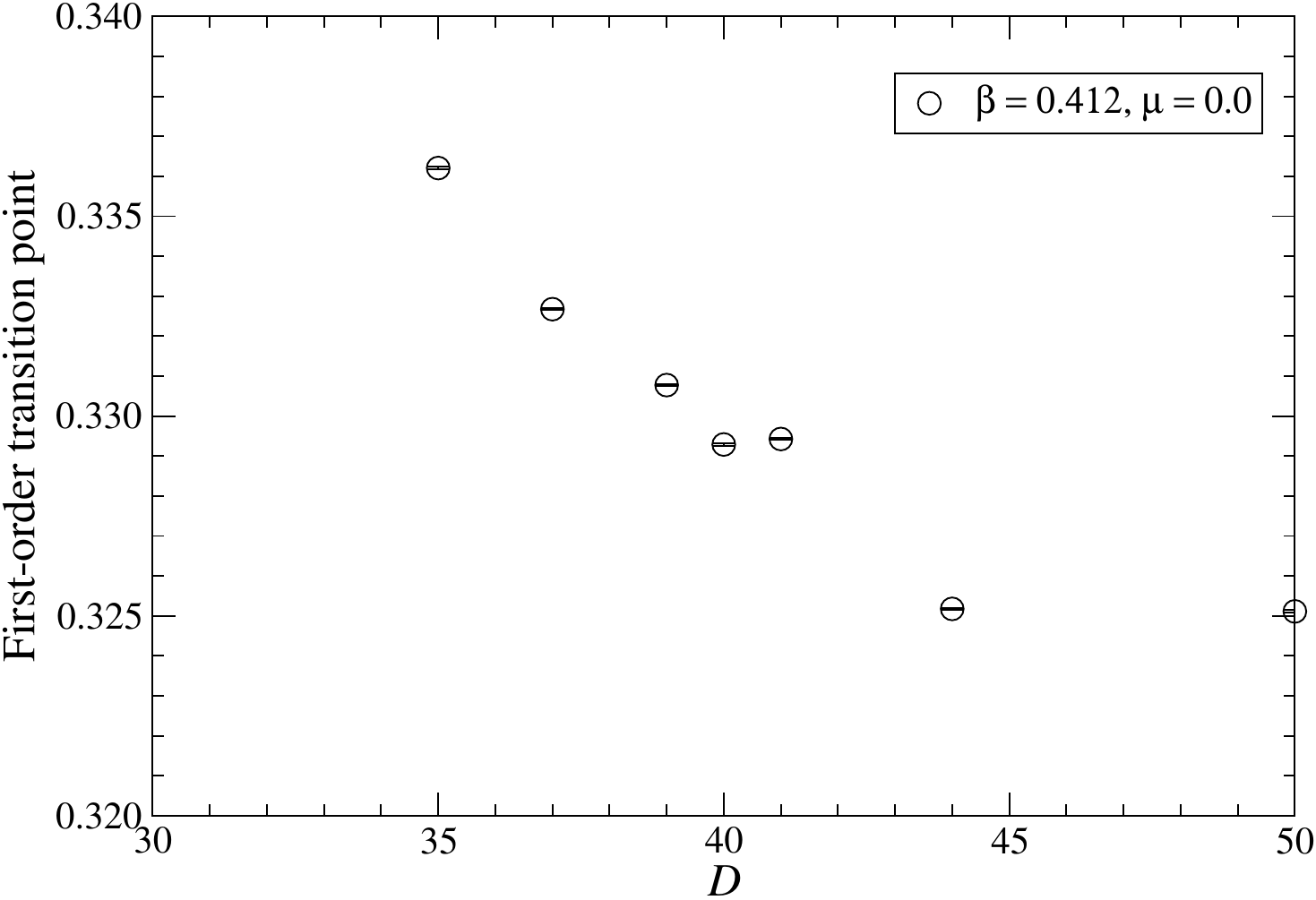}
	\caption{Convergence behavior of the transition point of $\eta$ as a function of $D$ at $\beta=0.412$ with vanishing $\mu$.}
  	\label{fig:eta_D}
\end{figure}

\begin{table}[htb]
	\caption{$\Delta \langle L\rangle$ and the first-order transition points of $(\beta,\eta)$ at $\mu=0$, $1$, $2$. All the results are obtained with $D=50$.}
	\label{tab:link_4d}
	\begin{center}
	  	\begin{tabular}{|c|c|c|}\hline
          	\multicolumn{3}{|c|}{$\mu=0$}  \\ \hline
			$\beta$ & $\eta$ & $\Delta \langle L\rangle$  \\ \hline
			0.412 & 0.32511(4) & 0.1448570626  \\
			0.414 & 0.32353(3) & 0.1746984961  \\
			0.415 & 0.32266(4) & 0.1930650060  \\
			0.416 & 0.32172(4) & 0.2096426007  \\
			0.417 & 0.32084(4) & 0.2156841414  \\ 
			0.418 & 0.32003(4) & 0.2358818267  \\ 
			0.419 & 0.31922(4) & 0.2434405732  \\ 
			0.420 & 0.31834(4) & 0.2578042490  \\ 
			0.425 & 0.31422(2) & 0.3007002064  \\ \hline
         		\multicolumn{3}{|c|}{$\mu=1$}  \\ \hline
			$\beta$ & $\eta$ & $\Delta \langle L\rangle$  \\ \hline
			0.415 & 0.28055(2) & 0.1443182714  \\
			0.416 & 0.27984(3) & 0.1815751895  \\
			0.417 & 0.27916(3) & 0.2059557920  \\
			0.418 & 0.27841(3) & 0.2232897426  \\
			0.419 & 0.27766(3) & 0.2468143002  \\
			0.420 & 0.27691(4) & 0.2614199552  \\
			0.421 & 0.27628(2) & 0.2735230025  \\ \hline
           		 \multicolumn{3}{|c|}{$\mu=2$}  \\ \hline
			$\beta$ & $\eta$ & $\Delta \langle L\rangle$  \\ \hline
			0.409 & 0.20972(3) & 0.0516254411  \\
			0.410 & 0.20894(3) & 0.1150573520  \\
			0.411 & 0.20816(4) & 0.1403103742  \\
			0.412 & 0.20741(3) & 0.1720190275  \\
			0.413 & 0.20661(3) & 0.1985852268  \\
			0.414 & 0.20584(3) & 0.2196185602  \\ \hline
	\end{tabular}
	\end{center}
\end{table}

\begin{figure}[htbp]
  	\centering
	\includegraphics[width=0.75\hsize]{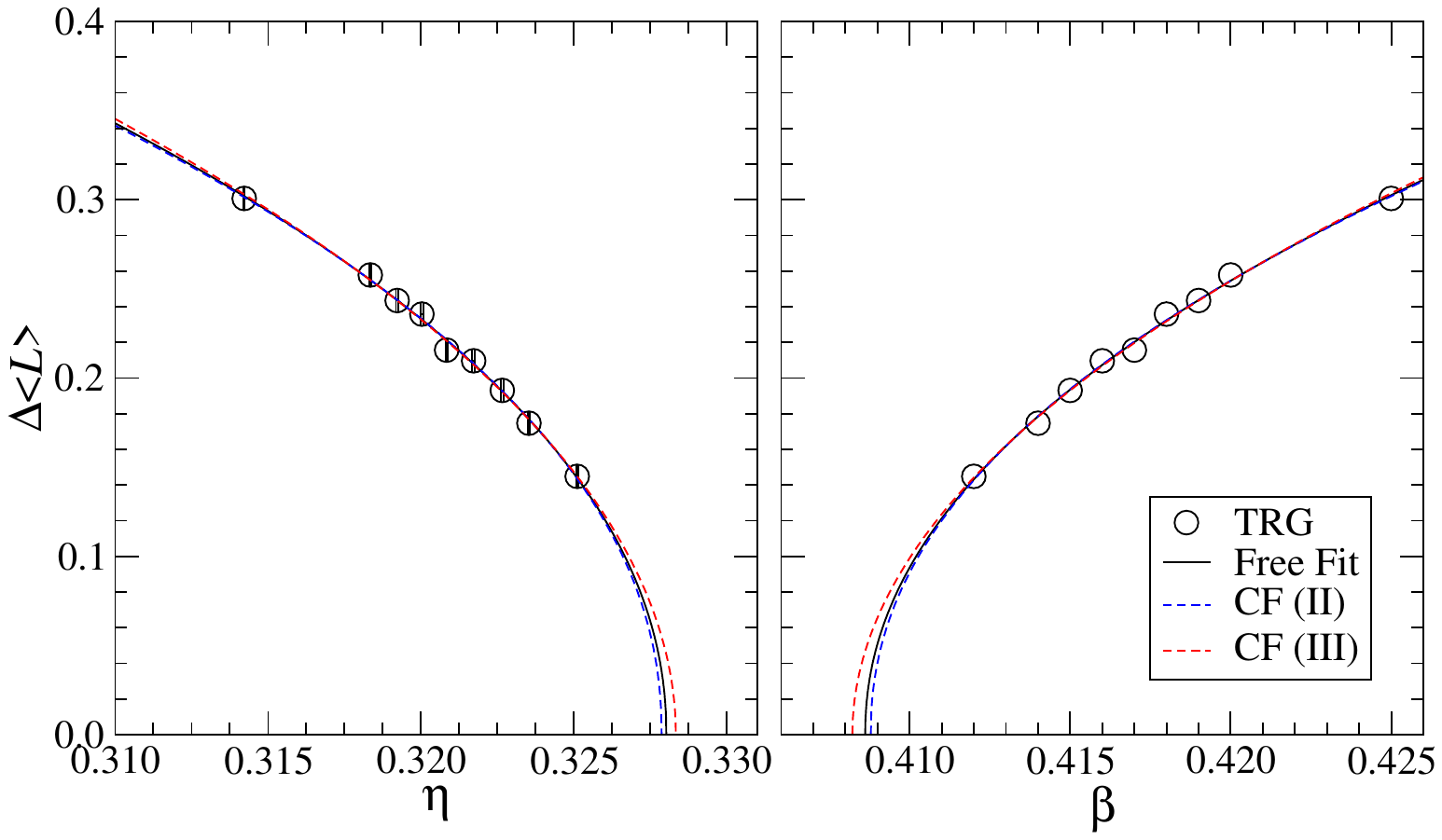}
	\caption{Fit of $\Delta \langle L\rangle$ at $\mu=0$ as a function of $\eta$ (left) and $\beta$ (right). Solid curve denotes the fit result with $(p,q)$, free and dotted curves are for constrained fits. See the text for the details.}
  	\label{fig:delta_4d_mu0_fit}
\end{figure}

\begin{table}[htb]
  \caption{Fit results for $\Delta \langle L\rangle$. All the results are obtained with the TRG method whose bond dimension is set to $D=50$. Difference between four types of fits gives an estimate of the systematic error. See the text for the details.}
	\label{tab:link_4d_fit}
	\begin{center}
	  \begin{tabular}{|c|ccc|ccc|}\hline
          \multicolumn{7}{|c|}{$\mu=0$}  \\ \hline
	Fit type & $A$ & $\beta_{\rm c}$ & $p$ & $B$ & $\eta_{\rm c}$ & $q$ \\ \hline
	Free &2.1(3) &  0.4086(6) & 0.47(4) & 2.3(4) & 0.3280(6) & 0.48(4) \\ 
	CF(II) & 2.0(2) & 0.4088(4) & 0.46(2) & 2.2(2) & 0.3279(4) & 0.46(3) \\ 
	CF(III) & 2.34(3) & 0.4082(2) & 0.5 & 2.55(3)    & 0.3283(2) & 0.5 \\ \hline
          \multicolumn{7}{|c|}{$\mu=1$}  \\ \hline
	Fit type & $A$ & $\beta_{\rm c}$ & $p$ & $B$ & $\eta_{\rm c}$ & $q$ \\ \hline
	Free & 1.5(2) &  0.4139(2) & 0.35(7) & 1.7(2) & 0.2813(2) & 0.34(3) \\ 
	CF(I) &2.5(4) & 0.4130(3) & 0.46(3) & 3.0(5)    & 0.2820(3) & 0.46(4) \\ 
	CF(II) & 2.6(3) & 0.4129(2) & 0.46(2) & 3.0(4)    & 0.2821(2) & 0.46(3) \\ 
	CF(III) & 3.04(6) & 0.4126(2) & 0.5 & 3.57(7)    & 0.2823(1) & 0.5 \\ \hline
          \multicolumn{7}{|c|}{$\mu=2$}  \\ \hline
	Fit type & $A$ & $\beta_{\rm c}$ & $p$ & $B$ & $\eta_{\rm c}$ & $q$ \\ \hline
	Free & 2.8(6) &  0.40873(7) & 0.48(4) & 3.2(8) & 0.20994(9) & 0.49(4) \\ 
	CF(I) & 2.4(4) &  0.40878(7) & 0.46(3) & 2.7(5) &  0.20990(8) & 0.46(4) \\ 
	CF(II) & 2.5(3) & 0.40877(5) & 0.46(2) & 2.8(4)  & 0.20990(6) & 0.46(3) \\ 
	CF(III) & 3.01(3) &  0.40869(3) & 0.5 & 3.42(5) &  0.20996(4) & 0.5 \\ \hline
	\end{tabular}
	\end{center}
\end{table}

Now, let us estimate the critical endpoint with the TRG approach.
The first-order phase transition line in the phase diagram terminates at the critical endpoint $(\beta_{\rm c},\eta_{\rm c})$. 
We employ the average link defined by
\begin{align}
\label{eq:def_al}
	\langle L\rangle=\frac{1}{4V}\frac{\partial \ln Z}{\partial(2\eta)}
\end{align}
to detect the first-order phase transition. 
We regard the critical endpoint $(\beta_{\rm c},\eta_{\rm c})$ as a point where the jump in $\langle L\rangle$, as a function of $\eta$, vanishes.
The factor $4V$ corresponds to the number of links in $\Lambda_{3+1}$ with the periodic boundary condition. 
We evaluate $\langle L\rangle$ with the impurity tensor method, whose expression is given in the appendix of Ref.~\cite{Akiyama:2022eip}.
In the following, all the results are calculated by setting $D=50$ in the thermodynamic limit, where the TRG computation converges for the system size. 
We first determine the critical endpoint in the $\mu=0$ case.
Figure~\ref{fig:link_4d_mu=0} shows the $\eta$ dependence of $\langle L\rangle$ at $\mu=0$ with the several choices of $\beta$. We observe clear gaps in $\langle L\rangle$ at a certain value of $\eta$ for $\beta\in[0.412,0.425]$.
Values of these gaps in $\langle L\rangle$, denoted by $\Delta \langle L\rangle$, are listed in Table~\ref{tab:link_4d}, together with the corresponding first-order transition point of $(\beta,\eta)$. 
Although $\Delta \langle L\rangle$ is evaluated just by $\langle L\rangle(\eta=\eta_{+})-\langle L\rangle(\eta=\eta_{-})$, where $\eta_{+}$ and $\eta_{-}$ are chosen from different phases, we set $\eta_{+}-\eta_{-}={\rm O}(10^{-5})$ for all $\beta$. The error for $\eta$ in Table~\ref{tab:link_4d} is provided by the magnitude of $\eta_{+}-\eta_{-}$. 
Figure~\ref{fig:eta_D} shows the typical $D$ dependence of the transition point $\eta$. 
The relative error between the first-order transition points with $D=44$ and $D=50$ is 0.019\%.
Hence, it may be expected that the finite-$D$ effect is well suppressed to identify transition points.
In order to determine the critical endpoint $(\beta_{\rm c},\eta_{\rm c})$, we separately fit the data of $\Delta \langle L\rangle$, assuming the functions $\Delta \langle L\rangle=A(\beta-\beta_{\rm c})^p$ and $\Delta \langle L\rangle=B(\eta_{\rm c}-\eta)^q$, respectively, where $A$, $B$, $\beta_{\rm c}$, $\eta_{\rm c}$, $p$, and $q$ are the fit parameters. 
The fit results are drawn in Fig.~\ref{fig:delta_4d_mu0_fit} and their numerical values are presented in Table~\ref{tab:link_4d_fit}. 
The fit provides us with $(\beta_{\rm c},\eta_{\rm c})=(0.409(7),0.3280(6))$ as the critical endpoint at $\mu=0$.

\begin{figure}[htbp]
	\centering
	\includegraphics[width=0.75\hsize]{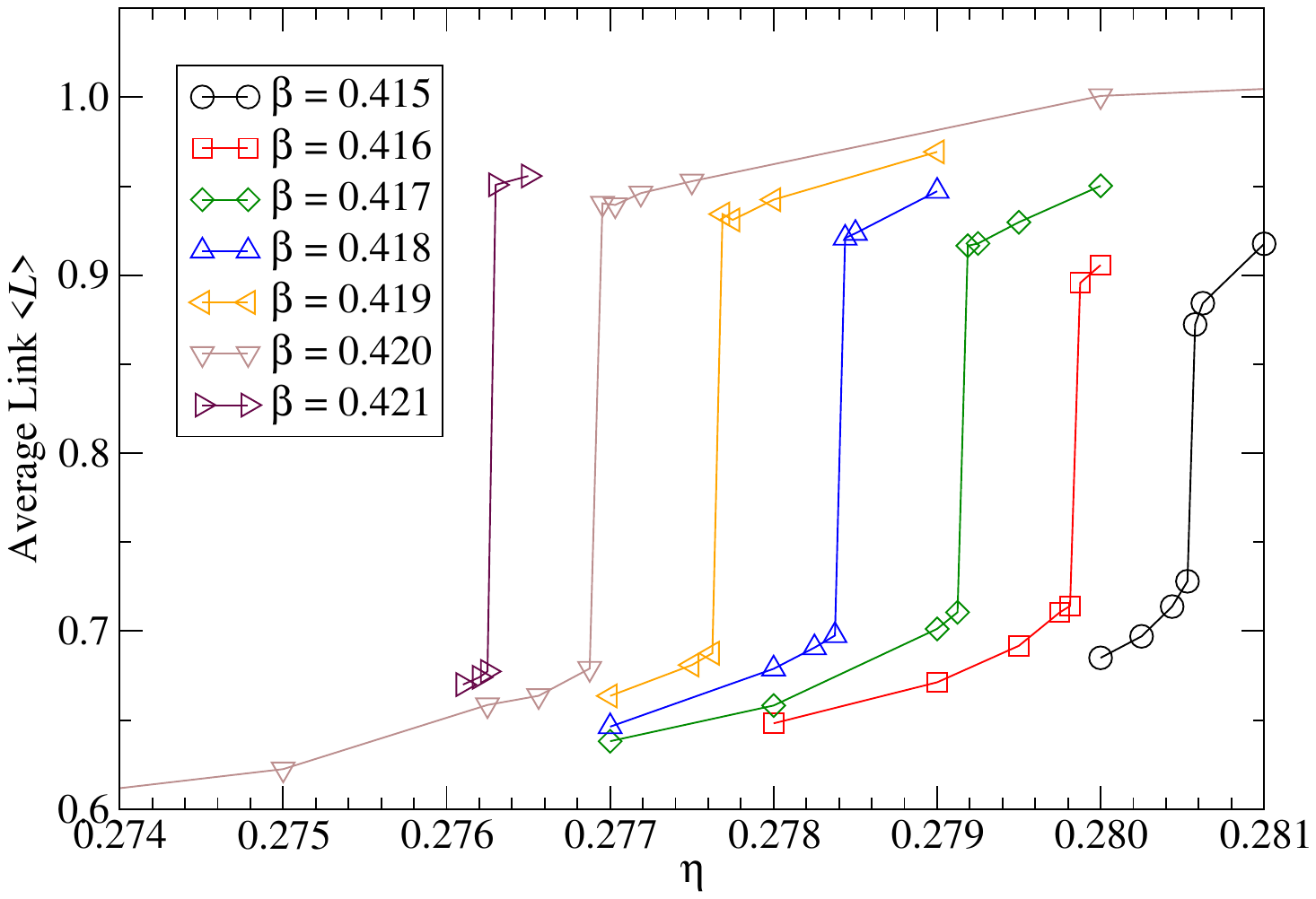}
	\caption{$\eta$ dependence of $\langle L\rangle$ at $\mu=1$ for $\beta\in [0.415,0.421]$.}
  	\label{fig:link_mu=1}
\end{figure}

\begin{figure}[htbp]
	\centering
	\includegraphics[width=0.75\hsize]{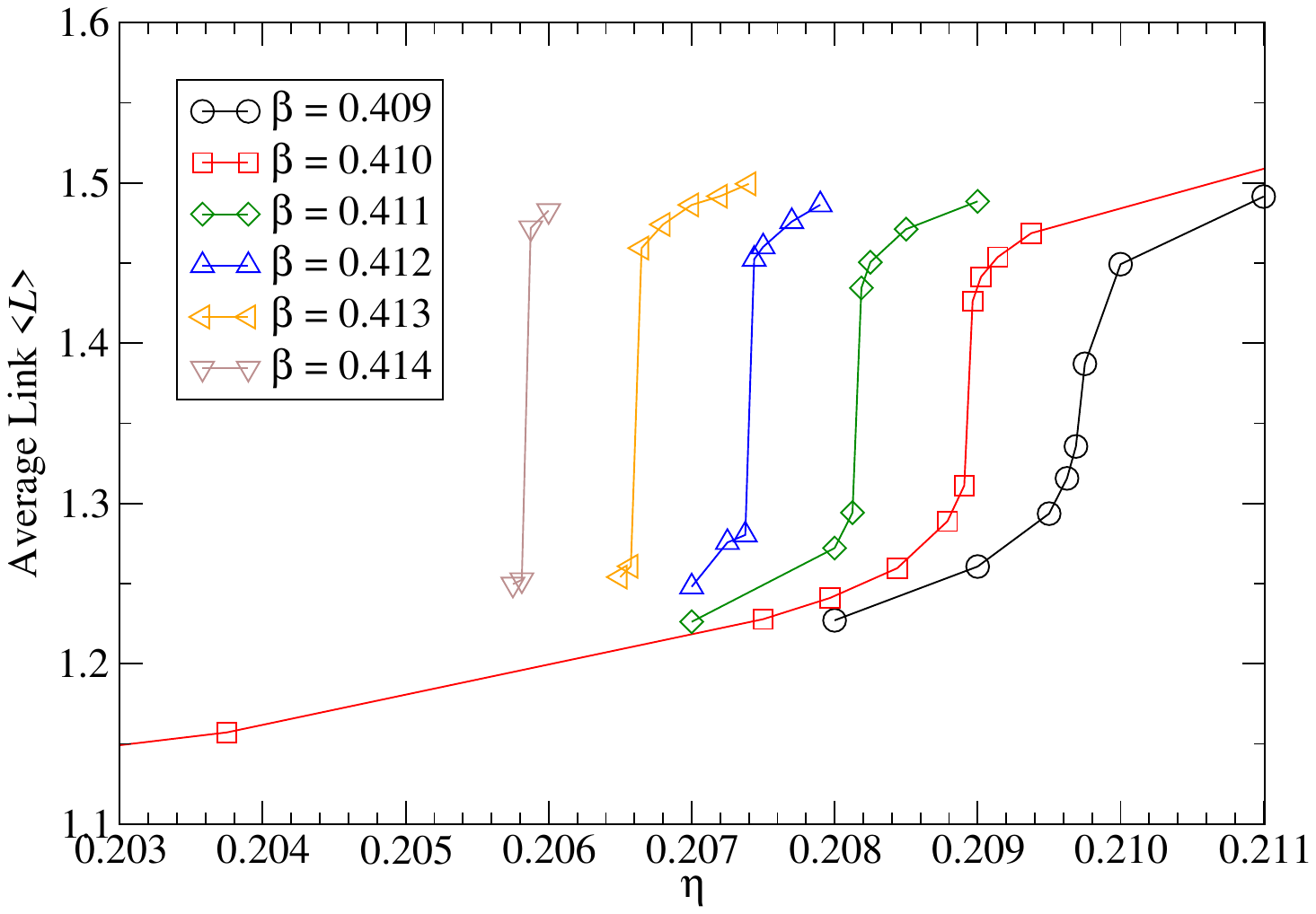}
	\caption{Same as Fig.~\ref{fig:link_mu=1} at $\mu=2$ for $\beta\in [0.409,0.414]$.}
  	\label{fig:link_mu=2}
\end{figure}

\begin{figure}[htbp]
  	\centering
	\includegraphics[width=0.75\hsize]{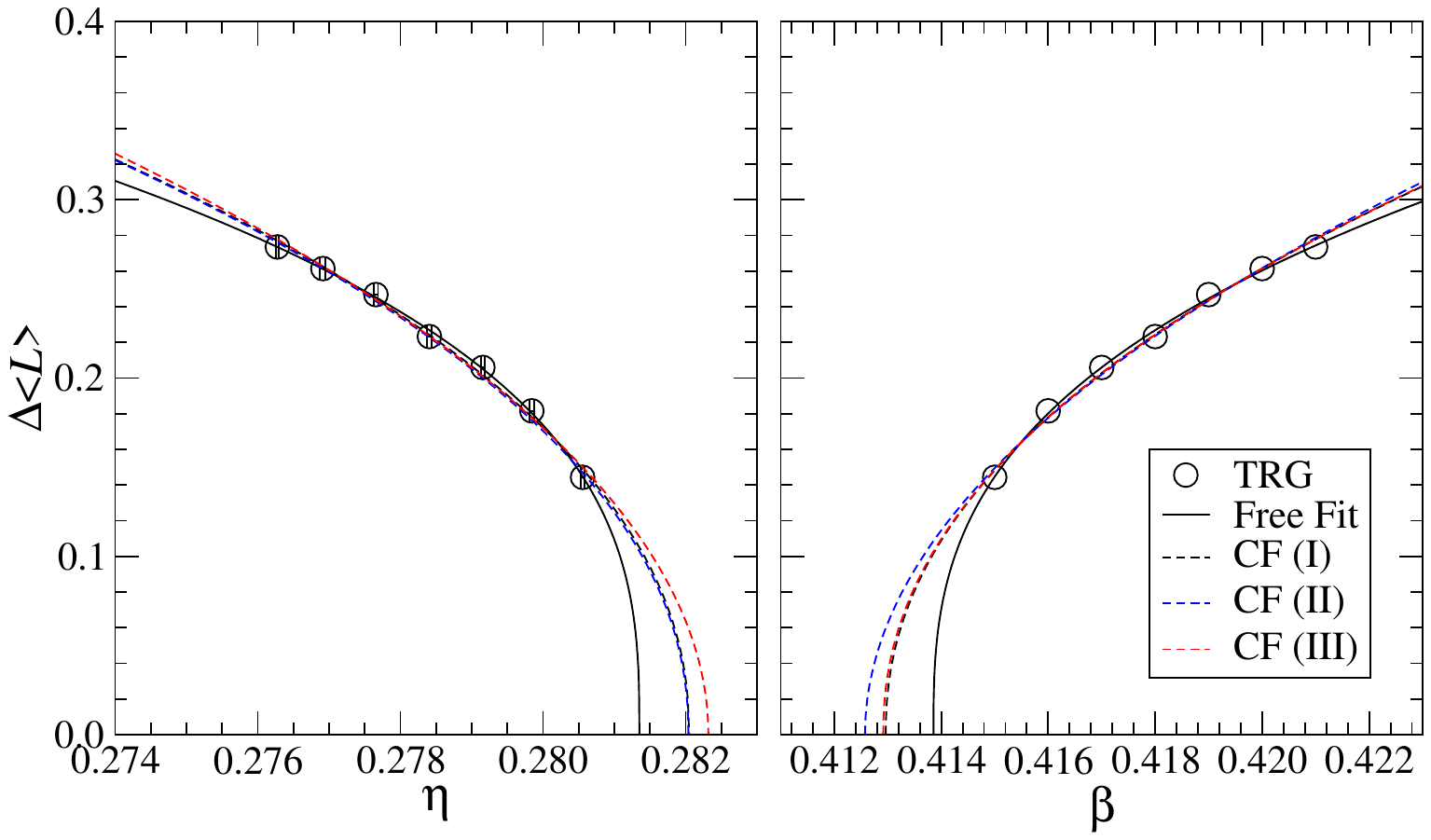}
	\caption{Fit of $\Delta \langle L\rangle$ at $\mu=1$ as a function of $\eta$ (left) and $\beta$ (right). Solid curve denotes the fit result with $(p,q)$ free, and dotted curves are for constrained fits. See the text for the details.
        }
  	\label{fig:delta_4d_mu1_fit}
\end{figure}

\begin{figure}[htbp]
  	\centering
	\includegraphics[width=0.75\hsize]{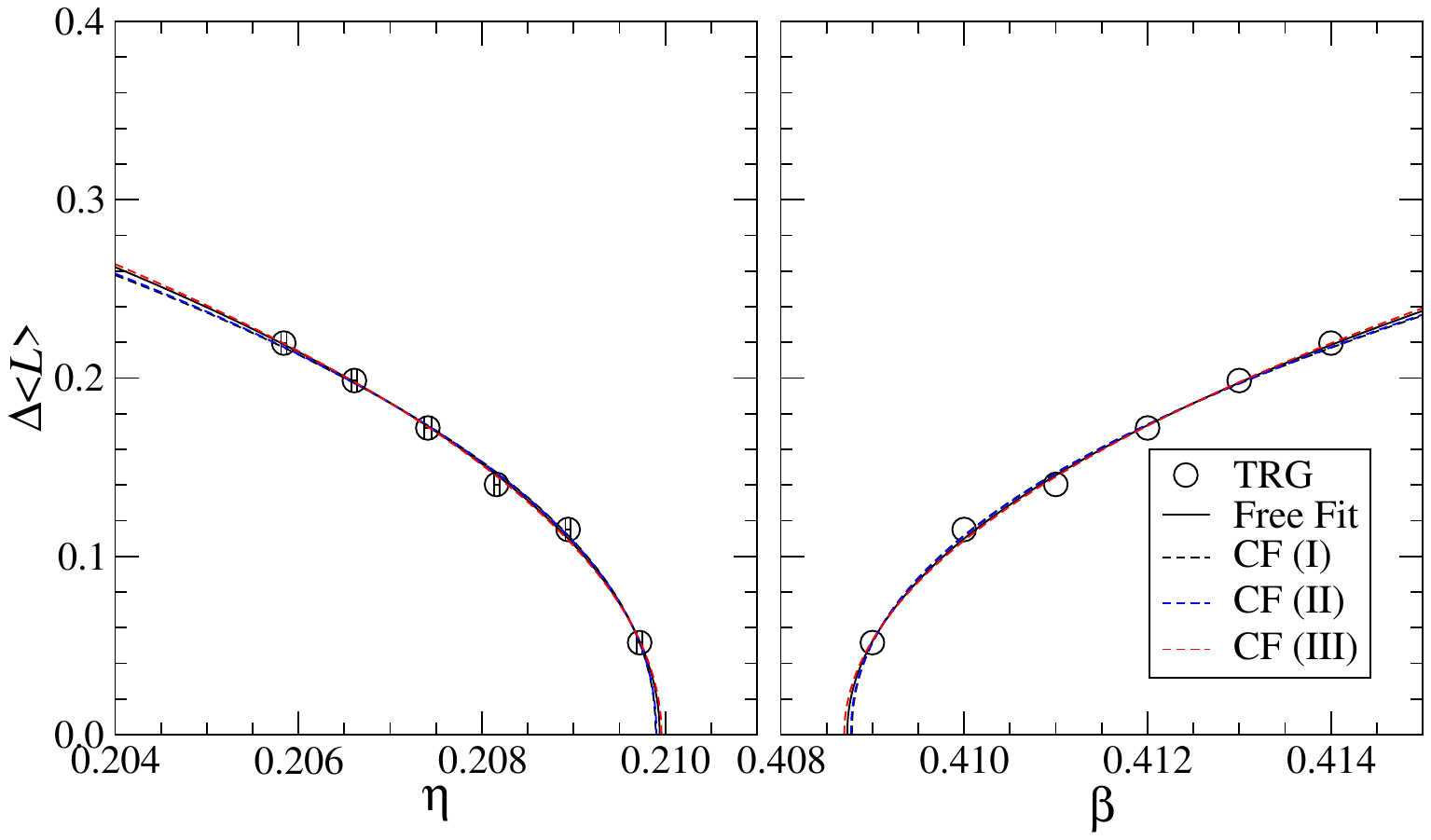}
	\caption{Same as Fig.~\ref{fig:delta_4d_mu1_fit} at $\mu=2$.}
  	\label{fig:delta_4d_mu2_fit}
\end{figure}

Let us turn to the finite density cases with $\mu=1$ and 2, where the Monte Carlo approach is ruled out by the sign problem. In Figs.~\ref{fig:link_mu=1} and \ref{fig:link_mu=2}  we plot the $\eta$ dependence of the link average with the several choices of $\beta$ at $\mu=1$ and 2, respectively. These should be compared with Figs.~10 and 11 of Ref.~\cite{Akiyama:2022eip} in the $\mathds{Z}_{2}$ model. 
We find a similar quality of data for both cases, which means that the TRG method works efficiently regardless of the sign problem. 
Table~\ref{tab:link_4d} summarizes the finite values of $\Delta \langle L\rangle$ and the transition points. $\Delta \langle L\rangle$ is fitted with the same functions as in the case of $\mu=0$. The fit results at $\mu=1$ and 2 are shown in Figs.~\ref{fig:delta_4d_mu1_fit} and \ref{fig:delta_4d_mu2_fit}, respectively. 
Their numerical values are presented in Table~\ref{tab:link_4d_fit}, together with the results of $\mu=0$.
We find that the values of $(p,q)$ at $\mu=1$ are relatively smaller than those at $\mu=0$ and 2 which are close to the mean-field values of $(p,q)=(0.5,0.5)$ expected from the conventional universality argument based on the dimensionality and the symmetry of the order parameter.
For an instructive purpose, we try three constrained fits.
The first one, which we call CF(I), is a simultaneous fit of the  $\mu=1$ and 2 data assuming $(p,q)$ in common, whose results are drawn by the black dotted curves in Figs.~\ref{fig:delta_4d_mu1_fit} and \ref{fig:delta_4d_mu2_fit} and numerical values are given in Table~\ref{tab:link_4d_fit}. In Fig.~\ref{fig:delta_4d_mu1_fit} we observe that the solid and dotted curves are almost degenerate in the range of the data points, while the extrapolated values of $(\beta_{\rm c},\eta_{\rm c})$ are deviated by about 0.2\%.  This situation indicates that the data with $\Delta \la L\ra < 0.1$ near the transition point is important to determine the critical exponents precisely. 
The reason why the values of $(p,q)$ at $\mu=1$ in the free fit are deviated from those at $\mu=0$ and 2 is that the data points at $\mu=1$ are not sufficiently close to the critical point compared to the $\mu=0$ and 2 cases. 
The second one called CF(II) is a simultaneous fit of the  $\mu=0$, $1$ and 2 data assuming $(p,q)$ in common. The blue dotted curves in Figs.~\ref{fig:delta_4d_mu0_fit}, \ref{fig:delta_4d_mu1_fit} and \ref{fig:delta_4d_mu2_fit} represent the fit results. We find little deviation of the blue dotted curve from the black one in Figs.~\ref{fig:delta_4d_mu1_fit} and \ref{fig:delta_4d_mu2_fit}. Numerical results in Table~\ref{tab:link_4d_fit} also show little difference between CF(I) and CF(II). The third one called CF(III) is a mean-field inspired fit with $(p,q)=(0.5,0.5)$ fixed. The fit results, which are depicted with the red dotted curves in Figs.~\ref{fig:delta_4d_mu0_fit}, \ref{fig:delta_4d_mu1_fit} and \ref{fig:delta_4d_mu2_fit} and numerically presented in Table~\ref{tab:link_4d_fit}, are quite similar to the CF(II) case. Taking account of the results for the four types of fits our estimate of the location of the critical endpoints is $(\beta_{\rm c},\eta_{\rm c})=(0.4086(6)(4),0.3280(6)(3))$, $(0.4139(2)(13),0.2813(2)(10))$ and $(0.40873(7)(5),0.20994(9)(4))$ at $\mu=0$, $1$, and $\mu=2$, respectively, where the second error denotes the systematic one due to the maximum difference between the free fit and the three constrained fits.

Comparing the critical endpoints at $\mu=0$, $1$, $2$, we find that $\beta_{\rm c}$ has little $\mu$ dependence, while $\eta_{\rm c}$ is sizably diminished as $\mu$ increases. 
Notice that similar behavior has been observed in the $\mathds{Z}_{2}$ model~\cite{Akiyama:2022eip}.
We summarize these results in Fig.~\ref{fig:phasedgm_mu}, where the critical endpoints in the $\mathds{Z}_{3}$ model is plotted together with those in the $\mathds{Z}_{2}$ model~\cite{Akiyama:2022eip}.  
According to Refs.~\cite{Creutz:1979he,Baig:1987ka}, the values of $\beta_{\rm c}$ and $\eta_{\rm c}$ are expected to increase as $N$ in $\mathds{Z}_{N}$ increases.
Therefore, the resulting endpoints by the TRG method seem reasonable for the increase of $N$.

\begin{figure}[htbp]
	\centering
	\includegraphics[width=0.75\hsize]{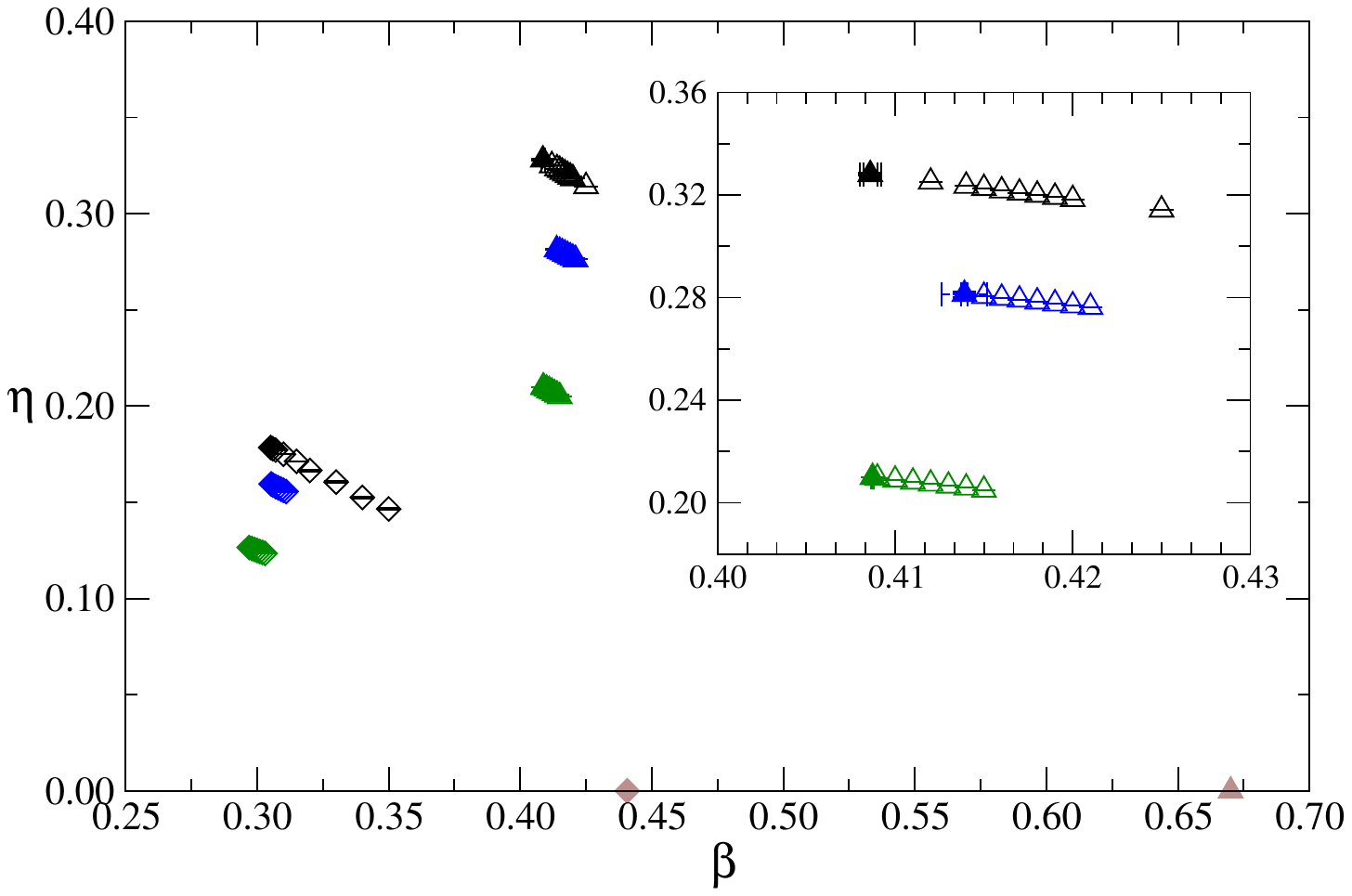}
	\caption{
	Summary of the critical points in the (3+1)$d$ $\mathds{Z}_{3}$ gauge-Higgs model, together with the $\mathds{Z}_{2}$ model~\cite{Akiyama:2022eip} for comparison. 
	Black, blue, and green symbols represent the transition points for $\mu=0$, $1$, and $2$, respectively. 
	Open symbols denote the first-order transition points. 
	Closed symbols are for the critical endpoints estimated by fitting.
	Diamonds and triangles denote the cases with $\mathds{Z}_{2}$ and $\mathds{Z}_{3}$, respectively.
	Symbols on the $\beta$-axis are the pure gauge transition points, which are $\ln(1+\sqrt{2})/2$ and $2\ln(1+\sqrt{3})/3$ for $\mathds{Z}_{2}$ and $\mathds{Z}_{3}$, respectively~\cite{Balian:1974ts,Balian:1974ir,Balian:1974xw,KorthalsAltes:1978tp,Yoneya:1978dt}.
	The inset graph makes a closer look for transition points of the $\mathds{Z}_{3}$ model including the estimated systematic errors.
	}
  	\label{fig:phasedgm_mu}
\end{figure}

\clearpage

\section{Summary and outlook} 
\label{sec:summary}

In this paper, we have determined the critical endpoints of (3+1)$d$ $\mathds{Z}_3$ gauge-Higgs model at finite density, where the sign problem prohibits the conventional Monte Carlo approach. 
We have closely followed the determination procedure employed in the previous work on the $\mathds{Z}_2$ model~\cite{Akiyama:2022eip}, which is free from the sign problem. 
The resulting endpoints by the TRG method seem reasonable, compared with the previous work of the $\mathds{Z}_2$ model~\cite{Akiyama:2022eip}.
In addition, we have computed the plaquette value as a function of $\beta$, which is comparable with the previous study of the same model by the dual lattice simulation~\cite{Gattringer:2012jt}.
Our results show that the TRG method works efficiently for both models regardless of the existence of the sign problem, so it is a promising approach to the future investigation of the QCD at finite density.
The next step should be an extension of this study to the (3+1)$d$ lattice gauge theories with continuous gauge groups, also including dynamical matter fields.

\begin{acknowledgments}
 Numerical calculation for the present work was carried out with Oakbridge-CX in the Information Technology Center at The University of Tokyo and the computational resource was offered under the category of General Projects by Research Institute for Information Technology in Kyushu University.
We also used the supercomputer Fugaku provided by RIKEN through the HPCI System Research Project (Project ID: hp210204, hp220203) and computational resources of Wisteria/BDEC-01 and Cygnus under the Multidisciplinary Cooperative Research Program of Center for Computational Sciences, University of Tsukuba. 
This work is supported in part by Grants-in-Aid for Scientific Research from the Ministry of Education, Culture, Sports, Science and Technology (MEXT) (No. 20H00148). 
S. A. is supported by the Endowed Project for Quantum Software Research and Education, the University of Tokyo (\cite{qsw}), and by JSPS KAKENHI Grant Number JP23K13096. 
\end{acknowledgments}


\appendix

\section{Impurity tensor method for $\langle U\rangle$}
\label{app:plaq}
 
In section~\ref{sec:results}, $\langle U\rangle$ is calculated by the impurity tensor method, where we slightly modify the asymmetric formulation in Ref.~\cite{Liu:2013nsa}.
To explain our modification, we first review the asymmetric formulation for (3+1)$d$ $\mathds{Z}_{3}$ gauge-Higgs model.
We regard the local Boltzmann weight corresponding to the plaquette interaction in Eq.~\eqref{eq:action} as a four-rank tensor,
\begin{align}
	&W_{U_{\nu}(n)U_{\rho}(n+\hat{\nu})U_{\nu}(n+\hat{\rho})U_{\rho}(n)}
	\nonumber\\
	&=
	\exp\left(
	\beta
	\Re\left[U_{\nu}(n)U_{\rho}(n+\hat{\nu})U^{*}_{\nu}(n+\hat{\rho})U^{*}_{\rho}(n)\right]
	\right)
	\nonumber\\
	&\times
	\exp\left[
	\frac{\eta}{3}
	\left\{
	\cosh\left(\mu\delta_{\nu,4}\right)\Re\left[U_{\nu}(n)\right]
	+{\rm i}\sinh\left(\mu\delta_{\nu,4}\right)\Im\left[U_{\nu}(n)\right]
	\right\},
	\right]
	\nonumber\\
	&\times
	\exp\left[
	\frac{\eta}{3}
	\left\{
	\cosh\left(\mu\delta_{\nu,4}\right)\Re\left[U_{\rho}(n+\hat{\nu})\right]
	+{\rm i}\sinh\left(\mu\delta_{\nu,4}\right)\Im\left[U_{\rho}(n+\hat{\nu})\right]
	\right\},
	\right]
	\nonumber\\
	&\times
	\exp\left[
	\frac{\eta}{3}
	\left\{
	\cosh\left(\mu\delta_{\nu,4}\right)\Re\left[U^{*}_{\nu}(n+\hat{\rho})\right]
	+{\rm i}\sinh\left(\mu\delta_{\nu,4}\right)\Im\left[U^{*}_{\nu}(n+\hat{\rho})\right]
	\right\},
	\right]
	\nonumber\\
	&\times
	\exp\left[
	\frac{\eta}{3}
	\left\{
	\cosh\left(\mu\delta_{\nu,4}\right)\Re\left[U^{*}_{\rho}(n)\right]
	+{\rm i}\sinh\left(\mu\delta_{\nu,4}\right)\Im\left[U^{*}_{\rho}(n)\right]
	\right\},
	\right]
	.
\end{align}
For later convenience, we have included spin-spin coupling terms associated with the plaquette.
Using the higher-order singular value decomposition (HOSVD), we can decompose $W$ via
\begin{align}
\label{eq:hosvd}
	W_{U_{\nu}(n)U_{\rho}(n+\hat{\nu})U_{\nu}(n+\hat{\rho})U_{\rho}(n)}
	=\sum_{a,b,c,d}V_{U_{\nu}(n)a}V_{U_{\rho}(n+\hat{\nu})b}V_{U_{\nu}(n+\hat{\rho})c}V_{U_{\rho}(n),d}B_{abcd},
\end{align}
where $V$'s are unitary matrices and $B$ is the core tensor. 
Integrating out all link variables $U_{\nu}(n)$'s in Eq.~\eqref{eq:Z_ug} at each link independently, we have a six-leg tensor $A$ at each link according to
\begin{align}
\label{eq:pure_link}
	A_{m_{1}m_{2}m_{3}m'_{1}m'_{2}m'_{3}}
	=\sum_{U_{\nu}(n)}
	\left(\prod_{i=1}^{3}V_{U_{\nu}(n)m_{i}}V_{U_{\nu}(n)m'_{i}}\right).
\end{align}
Therefore, we have the tensor network representation for Eq.~\eqref{eq:Z_ug} as
\begin{align}
\label{eq:asymmetric}
	Z={\rm tTr}\left[\left(\prod_{\ell}A_{\ell}\right)\left(\prod_{\square}B_{\square}\right)\right].
\end{align}
Considering the tensor contraction among $A_{\ell}$'s with $\ell=(n,1), (n,2), (n,3), (n,4)$ and $B_{\square}$'s associated with these $A_{\ell}$'s, we can define a local tensor $\mathcal{T}_{n}$ at the site $n$.
This is the asymmetric construction, which gives us the uniform tensor network representation such that
\begin{align}
\label{eq:tn_lgt}
	Z={\rm tTr}\left[\prod_{n\in\Lambda_{d+1}}\mathcal{T}_{n}\right]
	.
\end{align}
$\mathcal{T}_{n}$ can be regarded as an eight-leg tensor, say $\mathcal{T}_{n;xyztx'y'z't'}$, where each index is constructed by three indices coming from the HOSVD.

Now, we re-express $\mathcal{T}_{n}$ in Eq.~\eqref{eq:tn_lgt} using the original link variables $U_{\nu}(n)$, not using the indices introduced by the HOSVD. 
This can be easily done by re-expressing $A_{\ell}$'s in $\mathcal{T}_{n}$ as the right-hand side of Eq.~\eqref{eq:pure_link} and integrating out the indices coming from the HOSVD. 
One can see that this derivation gives us a new local tensor in the following form,
\begin{align}
\label{eq:plaq_tensor}
	&\mathcal{T}_{n;
	(x_{1}x_{2}x_{3})
	(y_{1}y_{2}y_{3})
	(z_{1}z_{2}z_{3})
	(t_{1}t_{2}t_{3})
	(x'_{1}x'_{2}x'_{3})
	(y'_{1}y'_{2}y'_{3})
	(z'_{1}z'_{2}z'_{3})
	(t'_{1}t'_{2}t'_{3})}
	=
	\nonumber\\
	&\left(\prod_{\nu=1}^{4}\sum_{U_{\nu}(n)}\right)
	\delta_{U_{1}(n)y_{1}z_{1}t_{1}}
	\delta_{U_{2}(n)x_{2}z_{2}t_{2}}
	\delta_{U_{3}(n)x_{3}y_{3}t_{3}}
	\delta_{U_{4}(n)x_{4}y_{4}z_{4}}
	\nonumber\\
	&\times
	W_{U_{1}(n)x'_{2}y'_{1}U_{2}(n)}
	W_{U_{1}(n)x'_{3}z'_{1}U_{3}(n)}
	W_{U_{1}(n)x'_{4}t'_{1}U_{4}(n)}
	W_{U_{2}(n)y'_{3}z'_{2}U_{3}(n)}
	W_{U_{2}(n)y'_{4}t'_{2}U_{4}(n)}
	W_{U_{3}(n)z'_{4}t'_{3}U_{4}(n)}
	.
\end{align}
Using this $\mathcal{T}_{n}$, the path integral is again represented as in Eq.~\eqref{eq:tn_lgt}.
Note that this construction has also been employed for 3$d$ $SU(2)$ gauge theory in Ref.~\cite{Kuwahara:2022ubg}.
Thanks to Eq.~\eqref{eq:plaq_tensor}, we can easily introduce the impurity tensor to describe $\langle U\rangle$.
For example, the expectation value of the plaquette on the $12$-plane is expressed by the following impurity tensor,
\begin{align}
	&\mathcal{S}^{[12]}_{n;
	(x_{1}x_{2}x_{3})
	(y_{1}y_{2}y_{3})
	(z_{1}z_{2}z_{3})
	(t_{1}t_{2}t_{3})
	(x'_{1}x'_{2}x'_{3})
	(y'_{1}y'_{2}y'_{3})
	(z'_{1}z'_{2}z'_{3})
	(t'_{1}t'_{2}t'_{3})}
	=
	\nonumber\\
	&\left(\prod_{\nu=1}^{4}\sum_{U_{\nu}(n)}\right)
	\delta_{U_{1}(n)y_{1}z_{1}t_{1}}
	\delta_{U_{2}(n)x_{2}z_{2}t_{2}}
	\delta_{U_{3}(n)x_{3}y_{3}t_{3}}
	\delta_{U_{4}(n)x_{4}y_{4}z_{4}}
	\exp\left(
	\beta
	\Re\left[U_{1}(n)x'_{2}(y')^{*}_{1}U^{*}_{2}(n)\right]
	\right)
	\nonumber\\
	&\times
	W_{U_{1}(n)x'_{2}y'_{1}U_{2}(n)}
	W_{U_{1}(n)x'_{3}z'_{1}U_{3}(n)}
	W_{U_{1}(n)x'_{4}t'_{1}U_{4}(n)}
	W_{U_{2}(n)y'_{3}z'_{2}U_{3}(n)}
	W_{U_{2}(n)y'_{4}t'_{2}U_{4}(n)}
	W_{U_{3}(n)z'_{4}t'_{3}U_{4}(n)}
	.
\end{align} 
Therefore, we can finally express  $\langle U\rangle$ as
\begin{align}
	 \langle U\rangle
	 =
	 \frac{1}{Z}{\rm tTr}\left[\mathcal{S}_{c}\prod_{n\neq c}\mathcal{T}_{n}\right]
	 ,
\end{align}
by introducing the following impurity tensor,
\begin{align}
	\mathcal{S}_{n}
	=\frac{1}{6}
	\left(
	\mathcal{S}^{[12]}_{n}
	+\mathcal{S}^{[13]}_{n}
	+\mathcal{S}^{[14]}_{n}
	+\mathcal{S}^{[23]}_{n}
	+\mathcal{S}^{[24]}_{n}
	+\mathcal{S}^{[34]}_{n}
	\right)
	.
\end{align}


\bibliographystyle{JHEP}
\bibliography{bib/formulation,bib/algorithm,bib/discrete,bib/grassmann,bib/continuous,bib/gauge,bib/review,bib/for_this_paper}

\providecommand{\href}[2]{#2}\begingroup\raggedright\begin{thebibliography}{10}

\bibitem{Levin:2006jai}
M.~Levin and C.~P. Nave, \emph{{Tensor renormalization group approach to
  two-dimensional classical lattice models}},
  \href{https://doi.org/10.1103/PhysRevLett.99.120601}{\emph{Phys. Rev. Lett.}
  {\bfseries 99} (2007) 120601},
  [\href{https://arxiv.org/abs/cond-mat/0611687}{{\ttfamily
  cond-mat/0611687}}].

\bibitem{Gu:2010yh}
Z.-C. Gu, F.~Verstraete and X.-G. Wen, \emph{{Grassmann tensor network states
  and its renormalization for strongly correlated fermionic and bosonic
  states}},  \href{https://arxiv.org/abs/1004.2563}{{\ttfamily 1004.2563}}.

\bibitem{PhysRevB.86.045139}
Z.~Y. Xie, J.~Chen, M.~P. Qin, J.~W. Zhu, L.~P. Yang and T.~Xiang,
  \emph{Coarse-graining renormalization by higher-order singular value
  decomposition}, \href{https://doi.org/10.1103/PhysRevB.86.045139}{\emph{Phys.
  Rev. B} {\bfseries 86} (Jul, 2012) 045139},
  [\href{https://arxiv.org/abs/1201.1144}{{\ttfamily 1201.1144}}].

\bibitem{Shimizu:2014uva}
Y.~Shimizu and Y.~Kuramashi, \emph{{Grassmann tensor renormalization group
  approach to one-flavor lattice Schwinger model}},
  \href{https://doi.org/10.1103/PhysRevD.90.014508}{\emph{Phys. Rev.}
  {\bfseries D90} (2014) 014508},
  [\href{https://arxiv.org/abs/1403.0642}{{\ttfamily 1403.0642}}].

\bibitem{Sakai:2017jwp}
R.~Sakai, S.~Takeda and Y.~Yoshimura, \emph{{Higher order tensor
  renormalization group for relativistic fermion systems}},
  \href{https://doi.org/10.1093/ptep/ptx080}{\emph{PTEP} {\bfseries 2017}
  (2017) 063B07}, [\href{https://arxiv.org/abs/1705.07764}{{\ttfamily
  1705.07764}}].

\bibitem{Adachi:2019paf}
D.~Adachi, T.~Okubo and S.~Todo, \emph{{Anisotropic Tensor Renormalization
  Group}}, \href{https://doi.org/10.1103/PhysRevB.102.054432}{\emph{Phys. Rev.
  B} {\bfseries 102} (2020) 054432},
  [\href{https://arxiv.org/abs/1906.02007}{{\ttfamily 1906.02007}}].

\bibitem{Kadoh:2019kqk}
D.~Kadoh and K.~Nakayama, \emph{{Renormalization group on a triad network}},
  \href{https://arxiv.org/abs/1912.02414}{{\ttfamily 1912.02414}}.

\bibitem{Akiyama:2020soe}
S.~Akiyama, Y.~Kuramashi, T.~Yamashita and Y.~Yoshimura, \emph{{Restoration of
  chiral symmetry in cold and dense Nambu--Jona-Lasinio model with tensor
  renormalization group}},
  \href{https://doi.org/10.1007/JHEP01(2021)121}{\emph{JHEP} {\bfseries 01}
  (2021) 121}, [\href{https://arxiv.org/abs/2009.11583}{{\ttfamily
  2009.11583}}].

\bibitem{PhysRevB.105.L060402}
D.~Adachi, T.~Okubo and S.~Todo, \emph{Bond-weighted tensor renormalization
  group}, \href{https://doi.org/10.1103/PhysRevB.105.L060402}{\emph{Phys. Rev.
  B} {\bfseries 105} (Feb, 2022) L060402},
  [\href{https://arxiv.org/abs/2011.01679}{{\ttfamily 2011.01679}}].

\bibitem{Kadoh:2021fri}
D.~Kadoh, H.~Oba and S.~Takeda, \emph{{Triad second renormalization group}},
  \href{https://doi.org/10.1007/JHEP04(2022)121}{\emph{JHEP} {\bfseries 04}
  (2022) 121}, [\href{https://arxiv.org/abs/2107.08769}{{\ttfamily
  2107.08769}}].

\bibitem{Banuls:2019rao}
M.~C. Ba{\~n}uls and K.~Cichy, \emph{{Review on Novel Methods for Lattice Gauge
  Theories}}, \href{https://doi.org/10.1088/1361-6633/ab6311}{\emph{Rept. Prog.
  Phys.} {\bfseries 83} (2020) 024401},
  [\href{https://arxiv.org/abs/1910.00257}{{\ttfamily 1910.00257}}].

\bibitem{Meurice:2020pxc}
Y.~Meurice, R.~Sakai and J.~Unmuth-Yockey, \emph{{Tensor lattice field theory
  for renormalization and quantum computing}},
  \href{https://doi.org/10.1103/RevModPhys.94.025005}{\emph{Rev. Mod. Phys.}
  {\bfseries 94} (2022) 025005},
  [\href{https://arxiv.org/abs/2010.06539}{{\ttfamily 2010.06539}}].

\bibitem{Okunishi:2021but}
K.~Okunishi, T.~Nishino and H.~Ueda, \emph{{Developments in the Tensor Network
  -- from Statistical Mechanics to Quantum Entanglement}},
  \href{https://doi.org/10.7566/JPSJ.91.062001}{\emph{J. Phys. Soc. Jap.}
  {\bfseries 91} (2022) 062001},
  [\href{https://arxiv.org/abs/2111.12223}{{\ttfamily 2111.12223}}].

\bibitem{Denbleyker:2013bea}
A.~Denbleyker, Y.~Liu, Y.~Meurice, M.~P. Qin, T.~Xiang, Z.~Y. Xie et~al.,
  \emph{{Controlling Sign Problems in Spin Models Using Tensor
  Renormalization}},
  \href{https://doi.org/10.1103/PhysRevD.89.016008}{\emph{Phys. Rev.}
  {\bfseries D89} (2014) 016008},
  [\href{https://arxiv.org/abs/1309.6623}{{\ttfamily 1309.6623}}].

\bibitem{Shimizu:2014fsa}
Y.~Shimizu and Y.~Kuramashi, \emph{{Critical behavior of the lattice Schwinger
  model with a topological term at $\theta=\pi$ using the Grassmann tensor
  renormalization group}},
  \href{https://doi.org/10.1103/PhysRevD.90.074503}{\emph{Phys. Rev.}
  {\bfseries D90} (2014) 074503},
  [\href{https://arxiv.org/abs/1408.0897}{{\ttfamily 1408.0897}}].

\bibitem{Takeda:2014vwa}
S.~Takeda and Y.~Yoshimura, \emph{{Grassmann tensor renormalization group for
  the one-flavor lattice Gross-Neveu model with finite chemical potential}},
  \href{https://doi.org/10.1093/ptep/ptv022}{\emph{PTEP} {\bfseries 2015}
  (2015) 043B01}, [\href{https://arxiv.org/abs/1412.7855}{{\ttfamily
  1412.7855}}].

\bibitem{Kawauchi:2016xng}
H.~Kawauchi and S.~Takeda, \emph{{Tensor renormalization group analysis of
  CP($N$-1) model}},
  \href{https://doi.org/10.1103/PhysRevD.93.114503}{\emph{Phys. Rev.}
  {\bfseries D93} (2016) 114503},
  [\href{https://arxiv.org/abs/1603.09455}{{\ttfamily 1603.09455}}].

\bibitem{Shimizu:2017onf}
Y.~Shimizu and Y.~Kuramashi, \emph{{Berezinskii-Kosterlitz-Thouless transition
  in lattice Schwinger model with one flavor of Wilson fermion}},
  \href{https://doi.org/10.1103/PhysRevD.97.034502}{\emph{Phys. Rev.}
  {\bfseries D97} (2018) 034502},
  [\href{https://arxiv.org/abs/1712.07808}{{\ttfamily 1712.07808}}].

\bibitem{Kadoh:2018hqq}
D.~Kadoh, Y.~Kuramashi, Y.~Nakamura, R.~Sakai, S.~Takeda and Y.~Yoshimura,
  \emph{{Tensor network formulation for two-dimensional lattice $ \mathcal{N} $
  = 1 Wess-Zumino model}},
  \href{https://doi.org/10.1007/JHEP03(2018)141}{\emph{JHEP} {\bfseries 03}
  (2018) 141}, [\href{https://arxiv.org/abs/1801.04183}{{\ttfamily
  1801.04183}}].

\bibitem{Kadoh:2019ube}
D.~Kadoh, Y.~Kuramashi, Y.~Nakamura, R.~Sakai, S.~Takeda and Y.~Yoshimura,
  \emph{{Investigation of complex $\phi^{4}$ theory at finite density in two
  dimensions using TRG}},
  \href{https://doi.org/10.1007/JHEP02(2020)161}{\emph{JHEP} {\bfseries 02}
  (2020) 161}, [\href{https://arxiv.org/abs/1912.13092}{{\ttfamily
  1912.13092}}].

\bibitem{Kuramashi:2019cgs}
Y.~Kuramashi and Y.~Yoshimura, \emph{{Tensor renormalization group study of
  two-dimensional U(1) lattice gauge theory with a $\theta$ term}},
  \href{https://doi.org/10.1007/JHEP04(2020)089}{\emph{JHEP} {\bfseries 04}
  (2020) 089}, [\href{https://arxiv.org/abs/1911.06480}{{\ttfamily
  1911.06480}}].

\bibitem{Butt:2019uul}
N.~Butt, S.~Catterall, Y.~Meurice, R.~Sakai and J.~Unmuth-Yockey, \emph{{Tensor
  network formulation of the massless Schwinger model with staggered
  fermions}}, \href{https://doi.org/10.1103/PhysRevD.101.094509}{\emph{Phys.
  Rev. D} {\bfseries 101} (2020) 094509},
  [\href{https://arxiv.org/abs/1911.01285}{{\ttfamily 1911.01285}}].

\bibitem{Takeda:2021mnc}
S.~Takeda, \emph{{A novel method to evaluate real-time path integral for scalar
  $\phi^4$ theory}},  8, 2021,
  \href{https://arxiv.org/abs/2108.10017}{{\ttfamily 2108.10017}}.

\bibitem{Nakayama:2021iyp}
K.~Nakayama, L.~Funcke, K.~Jansen, Y.-J. Kao and S.~K\"uhn, \emph{{Phase
  structure of the CP(1) model in the presence of a topological
  \ensuremath{\theta}-term}},
  \href{https://doi.org/10.1103/PhysRevD.105.054507}{\emph{Phys. Rev. D}
  {\bfseries 105} (2022) 054507},
  [\href{https://arxiv.org/abs/2107.14220}{{\ttfamily 2107.14220}}].

\bibitem{Yoshimura:2017jpk}
Y.~Yoshimura, Y.~Kuramashi, Y.~Nakamura, S.~Takeda and R.~Sakai,
  \emph{{Calculation of fermionic Green functions with Grassmann higher-order
  tensor renormalization group}},
  \href{https://doi.org/10.1103/PhysRevD.97.054511}{\emph{Phys. Rev.}
  {\bfseries D97} (2018) 054511},
  [\href{https://arxiv.org/abs/1711.08121}{{\ttfamily 1711.08121}}].

\bibitem{Akiyama:2021xxr}
S.~Akiyama and Y.~Kuramashi, \emph{{Tensor renormalization group approach to
  (1+1)-dimensional Hubbard model}},
  \href{https://doi.org/10.1103/PhysRevD.104.014504}{\emph{Phys. Rev. D}
  {\bfseries 104} (2021) 014504},
  [\href{https://arxiv.org/abs/2105.00372}{{\ttfamily 2105.00372}}].

\bibitem{Bloch:2022vqz}
J.~Bloch and R.~Lohmayer, \emph{{Grassmann higher-order tensor renormalization
  group approach for two-dimensional strong-coupling QCD}},
  \href{https://doi.org/10.1016/j.nuclphysb.2022.116032}{\emph{Nucl. Phys. B}
  {\bfseries 986} (2023) 116032},
  [\href{https://arxiv.org/abs/2206.00545}{{\ttfamily 2206.00545}}].

\bibitem{Akiyama:2019xzy}
S.~Akiyama, Y.~Kuramashi, T.~Yamashita and Y.~Yoshimura, \emph{{Phase
  transition of four-dimensional Ising model with higher-order tensor
  renormalization group}},
  \href{https://doi.org/10.1103/PhysRevD.100.054510}{\emph{Phys. Rev.}
  {\bfseries D100} (2019) 054510},
  [\href{https://arxiv.org/abs/1906.06060}{{\ttfamily 1906.06060}}].

\bibitem{Akiyama:2020ntf}
S.~Akiyama, D.~Kadoh, Y.~Kuramashi, T.~Yamashita and Y.~Yoshimura,
  \emph{{Tensor renormalization group approach to four-dimensional complex
  $\phi^4$ theory at finite density}},
  \href{https://doi.org/10.1007/JHEP09(2020)177}{\emph{JHEP} {\bfseries 09}
  (2020) 177}, [\href{https://arxiv.org/abs/2005.04645}{{\ttfamily
  2005.04645}}].

\bibitem{Akiyama:2021zhf}
S.~Akiyama, Y.~Kuramashi and Y.~Yoshimura, \emph{{Phase transition of
  four-dimensional lattice $\phi^4$ theory with tensor renormalization group}},
  \href{https://doi.org/10.1103/PhysRevD.104.034507}{\emph{Phys. Rev. D}
  {\bfseries 104} (2021) 034507},
  [\href{https://arxiv.org/abs/2101.06953}{{\ttfamily 2101.06953}}].

\bibitem{Milde:2021vln}
P.~Milde, J.~Bloch and R.~Lohmayer, \emph{{Tensor-network simulation of the
  strong-coupling $U(N)$ model}},  12, 2021,
  \href{https://arxiv.org/abs/2112.01906}{{\ttfamily 2112.01906}}.

\bibitem{Akiyama:2022eip}
S.~Akiyama and Y.~Kuramashi, \emph{{Tensor renormalization group study of
  (3+1)-dimensional \ensuremath{\mathbb{Z}}$_{2}$ gauge-Higgs model at finite
  density}}, \href{https://doi.org/10.1007/JHEP05(2022)102}{\emph{JHEP}
  {\bfseries 05} (2022) 102},
  [\href{https://arxiv.org/abs/2202.10051}{{\ttfamily 2202.10051}}].

\bibitem{Unmuth-Yockey:2018ugm}
J.~Unmuth-Yockey, J.~Zhang, A.~Bazavov, Y.~Meurice and S.-W. Tsai,
  \emph{{Universal features of the Abelian Polyakov loop in 1+1 dimensions}},
  \href{https://doi.org/10.1103/PhysRevD.98.094511}{\emph{Phys. Rev.}
  {\bfseries D98} (2018) 094511},
  [\href{https://arxiv.org/abs/1807.09186}{{\ttfamily 1807.09186}}].

\bibitem{Bazavov:2019qih}
A.~Bazavov, S.~Catterall, R.~G. Jha and J.~Unmuth-Yockey, \emph{{Tensor
  renormalization group study of the non-Abelian Higgs model in two
  dimensions}}, \href{https://doi.org/10.1103/PhysRevD.99.114507}{\emph{Phys.
  Rev.} {\bfseries D99} (2019) 114507},
  [\href{https://arxiv.org/abs/1901.11443}{{\ttfamily 1901.11443}}].

\bibitem{Fukuma:2021cni}
M.~Fukuma, D.~Kadoh and N.~Matsumoto, \emph{{Tensor network approach to
  two-dimensional Yang\textendash{}Mills theories}},
  \href{https://doi.org/10.1093/ptep/ptab143}{\emph{PTEP} {\bfseries 2021}
  (2021) 123B03}, [\href{https://arxiv.org/abs/2107.14149}{{\ttfamily
  2107.14149}}].

\bibitem{Hirasawa:2021qvh}
M.~Hirasawa, A.~Matsumoto, J.~Nishimura and A.~Yosprakob, \emph{{Tensor
  renormalization group and the volume independence in 2D U(N) and SU(N) gauge
  theories}}, \href{https://doi.org/10.1007/JHEP12(2021)011}{\emph{JHEP}
  {\bfseries 12} (2021) 011},
  [\href{https://arxiv.org/abs/2110.05800}{{\ttfamily 2110.05800}}].

\bibitem{Dittrich:2014mxa}
B.~Dittrich, S.~Mizera and S.~Steinhaus, \emph{{Decorated tensor network
  renormalization for lattice gauge theories and spin foam models}},
  \href{https://doi.org/10.1088/1367-2630/18/5/053009}{\emph{New J. Phys.}
  {\bfseries 18} (2016) 053009},
  [\href{https://arxiv.org/abs/1409.2407}{{\ttfamily 1409.2407}}].

\bibitem{Kuramashi:2018mmi}
Y.~Kuramashi and Y.~Yoshimura, \emph{{Three-dimensional finite temperature
  Z$_{2}$ gauge theory with tensor network scheme}},
  \href{https://doi.org/10.1007/JHEP08(2019)023}{\emph{JHEP} {\bfseries 08}
  (2019) 023}, [\href{https://arxiv.org/abs/1808.08025}{{\ttfamily
  1808.08025}}].

\bibitem{Unmuth-Yockey:2018xak}
J.~F. Unmuth-Yockey, \emph{{Gauge-invariant rotor Hamiltonian from dual
  variables of 3D $U(1)$ gauge theory}},
  \href{https://doi.org/10.1103/PhysRevD.99.074502}{\emph{Phys. Rev. D}
  {\bfseries 99} (2019) 074502},
  [\href{https://arxiv.org/abs/1811.05884}{{\ttfamily 1811.05884}}].

\bibitem{Kuwahara:2022ubg}
T.~Kuwahara and A.~Tsuchiya, \emph{{Toward tensor renormalization group study
  of three-dimensional non-Abelian gauge theory}},
  \href{https://doi.org/10.1093/ptep/ptac103}{\emph{PTEP} {\bfseries 2022}
  (2022) 093B02}, [\href{https://arxiv.org/abs/2205.08883}{{\ttfamily
  2205.08883}}].

\bibitem{Creutz:1979he}
M.~Creutz, \emph{{Phase Diagrams for Coupled Spin Gauge Systems}},
  \href{https://doi.org/10.1103/PhysRevD.21.1006}{\emph{Phys. Rev. D}
  {\bfseries 21} (1980) 1006}.

\bibitem{Creutz:1983ev}
M.~Creutz, L.~Jacobs and C.~Rebbi, \emph{{Monte Carlo Computations in Lattice
  Gauge Theories}},
  \href{https://doi.org/10.1016/0370-1573(83)90016-9}{\emph{Phys. Rept.}
  {\bfseries 95} (1983) 201--282}.

\bibitem{Baig:1987ka}
M.~Baig, \emph{{Determination of the phase structure of the four-dimensional
  coupled gauge-Higgs Potts model}},
  \href{https://doi.org/10.1016/0370-2693(88)90579-5}{\emph{Phys. Lett. B}
  {\bfseries 207} (1988) 300--304}.

\bibitem{Gattringer:2012jt}
C.~Gattringer and A.~Schmidt, \emph{{Gauge and matter fields as surfaces and
  loops - an exploratory lattice study of the $Z_{3}$ Gauge-Higgs model}},
  \href{https://doi.org/10.1103/PhysRevD.86.094506}{\emph{Phys. Rev. D}
  {\bfseries 86} (2012) 094506},
  [\href{https://arxiv.org/abs/1208.6472}{{\ttfamily 1208.6472}}].

\bibitem{Langfeld:2018ykv}
K.~Langfeld, \emph{{String-like theory as solution to the sign problem of a
  finite density gauge theory}},
  \href{https://doi.org/10.22323/1.336.0049}{\emph{PoS} {\bfseries
  Confinement2018} (2018) 049},
  [\href{https://arxiv.org/abs/1811.12921}{{\ttfamily 1811.12921}}].

\bibitem{Balian:1974ts}
R.~Balian, J.~M. Drouffe and C.~Itzykson, \emph{{Gauge Fields on a Lattice. 1.
  General Outlook}},
  \href{https://doi.org/10.1103/PhysRevD.10.3376}{\emph{Phys. Rev. D}
  {\bfseries 10} (1974) 3376}.

\bibitem{Balian:1974ir}
R.~Balian, J.~M. Drouffe and C.~Itzykson, \emph{{Gauge Fields on a Lattice. 2.
  Gauge Invariant Ising Model}},
  \href{https://doi.org/10.1103/PhysRevD.11.2098}{\emph{Phys. Rev. D}
  {\bfseries 11} (1975) 2098}.

\bibitem{Balian:1974xw}
R.~Balian, J.~M. Drouffe and C.~Itzykson, \emph{{Gauge Fields on a Lattice. 3.
  Strong Coupling Expansions and Transition Points}},
  \href{https://doi.org/10.1103/PhysRevD.11.2104}{\emph{Phys. Rev. D}
  {\bfseries 11} (1975) 2104}.

\bibitem{KorthalsAltes:1978tp}
C.~P. Korthals~Altes, \emph{{Duality for $Z(N$) Gauge Theories}},
  \href{https://doi.org/10.1016/0550-3213(78)90207-9}{\emph{Nucl. Phys. B}
  {\bfseries 142} (1978) 315--326}.

\bibitem{Yoneya:1978dt}
T.~Yoneya, \emph{{$Z(N$) Topological Excitations in {Yang-Mills} Theories:
  Duality and Confinement}},
  \href{https://doi.org/10.1016/0550-3213(78)90502-3}{\emph{Nucl. Phys. B}
  {\bfseries 144} (1978) 195--218}.

\bibitem{PhysRevLett.43.799}
H.~W.~J. Bl\"ote and R.~H. Swendsen, \emph{First-order phase transitions and
  the three-state potts model},
  \href{https://doi.org/10.1103/PhysRevLett.43.799}{\emph{Phys. Rev. Lett.}
  {\bfseries 43} (Sep, 1979) 799--802}.

\bibitem{Caracciolo:1986ik}
S.~Caracciolo, G.~Parisi and S.~Patarnello, \emph{{Phase diagram for a
  ferromagnetic system with Potts symmetry in four-dimensions}},
  \href{https://doi.org/10.1209/0295-5075/4/1/002}{\emph{EPL} {\bfseries 4}
  (1987) 7--14}.

\bibitem{Liu:2013nsa}
Y.~Liu, Y.~Meurice, M.~P. Qin, J.~Unmuth-Yockey, T.~Xiang, Z.~Y. Xie et~al.,
  \emph{{Exact Blocking Formulas for Spin and Gauge Models}},
  \href{https://doi.org/10.1103/PhysRevD.88.056005}{\emph{Phys. Rev.}
  {\bfseries D88} (2013) 056005},
  [\href{https://arxiv.org/abs/1307.6543}{{\ttfamily 1307.6543}}].

\bibitem{qsw}
\url{https://qsw.phys.s.u-tokyo.ac.jp/}.

\end{thebibliography}\endgroup

\end{document}